\documentclass{aa}  

\usepackage{graphicx}
\usepackage{txfonts}
\begin{document}



    \title{The role of plasma $\beta$ in global coronal models}

   \subtitle{Bringing balance back to the force}

   \author{M. Brchnelova
          \inst{1}
          \and
          B. Ku\'zma\inst{2} \inst{3} \and
          F. Zhang \inst{1} \and
          A. Lani \inst{1}
          \and 
          S. Poedts \inst{1} \inst{4}
          }

   \institute{Centre for Mathematical Plasma Astrophysics, KU Leuven,
              Celestijnenlaan 200B, 3001, Leuven\\
              \email{michaela.brchnelova@kuleuven.be}
         \and
             Shenzhen Key Laboratory of Numerical Prediction for Space Storm, Institute of Space Science and Applied Technology, Harbin Institute of Technology, Shenzhen, People’s Republic of China, 518055
        \and 
             Key Laboratory of Solar Activity and Space Weather, National Space Science Center, Chinese Academy of Sciences, Beijing, People's Republic of China, 100190
        \and Institute of Physics, University of Maria Curie-Sk{\l}odowska, ul.\ Radziszewskiego 10, 20-031 Lublin, Poland 
             }
   \date{}

 
  \abstract
   {COolfluid COrona uNstrUcTured (COCONUT) is a global coronal magnetohydrodynamic (MHD) model that has been recently developed and will soon be integrated into the ESA Virtual Space Weather Modelling Centre (VSWMC). In order to achieve robustness and fast convergence to steady-state for numerical simulations with COCONUT, several assumptions and simplifications have been made during its development, such as prescribing filtered photospheric magnetic maps for representing the magnetic field conditions in the lower corona. This filtering leads to smoothing and lower magnetic field values at the inner boundary (i.e., the solar surface), resulting in an unrealistically high plasma $\beta$ (more than 1 in a large portion of the domain).}
   {In this paper, we examine the effects of prescribing such filtered/smoothed magnetograms in global coronal simulations and formulate a method for achieving more realistic plasma $\beta$ values and improving the resolution of electromagnetic features without losing computational performance.}
   {We make use of the newly developed COCONUT solver to demonstrate the effects of the highly pre-processed magnetic maps set at the inner boundary and the resulting high plasma $\beta$ on the features in the computational domain. Then, in our new approach, we shift the inner boundary to $2\;R_\oplus$ from the original $1.01\;R_\oplus$ and preserve the prescribed highly filtered magnetic map. With the shifted boundary, naturally, the boundary density and pressure are also adjusted to better represent the considered physical location. This effectively reduces the prescribed plasma $\beta$ and leads to a more realistic setup. The method is applied on a magnetic dipole, a minimum (2008) and a maximum (2012) solar activity case, to demonstrate its effects.}
   {The results obtained with the proposed approach show significant improvements in the resolved density and radial velocity profiles, and far more realistic values of the plasma $\beta$ at the boundary and inside the computational domain. This is also demonstrated via synthetic white light imaging (WLI) and with the validation against tomography data. The computational performance comparison shows similar convergence to a limit residual on the same grid when compared to the original setup. Considering that the grid can be further coarsened with this new setup, as its capacity to resolve features or structures is superior, the operational performance can be additionally increased if needed.}
   {The newly developed method is thus deemed as a good potential replacement of the original setup for operational purposes, providing higher physical detail of the resolved profiles while preserving a good convergence and robustness of the solver.}

   \keywords{ Magnetohydrodynamics (MHD) --
                Sun: corona --
                Methods: numerical
               }

   \maketitle
%
\section{Introduction}

Space weather forecasting has been gaining importance in the past few decades as our society is relying more and more on digital technologies and space infrastructure, both of which are fairly susceptible to space weather effects. Space weather modelling is, however, not a straightforward task due to the different environments that the solar wind and transients pass through on their way from the Sun's to the Earth's atmosphere. Typically, these different environments require different physics to be considered, and this is why software frameworks such as the ESA Virtual Space Weather Modelling Centre (VSWMC) \cite{vswmc},have been developed, allowing for coupling a chain of models to each other for forecasting purposes. The current heliospheric wind and CME evolution modelling in the VSWMC is heavily relying on the Wang-Sheeley-Arge (WSA,  \citet{Arge2003}) like coronal model in EUHFORIA \citep{Pomoell18}, because the alternative coronal models (Wind-Predict \citep{Reville2015} and Multi-VP \citep{Samara2021}) require substantially more CPU time. The WSA model is, however, semi-empirical and has been repeatedly shown to produce inaccurate boundary data at 0.1~AU\cite{Samara2021}. Yet, the coronal model is of great importance to space weather predictions since it is the first model on the Sun side, propagating the plasma field to the rest of the chain, e.g.\ to the EUHFORIA Heliosphere module \citep{Pomoell18} and the faster (AMR) version ICARUS \citep{Verbeke19}. Thus, any errors generated by this coronal model will spoil the prediction, regardless of the accuracy of the models that follow. 

In previous work (see \citet{PerriLeitner2022}) we have introduced COCONUT, a \emph{global coronal modelling} tool, relying on ideal-MHD, which can alternatively be used to predict the plasma properties at 0.1~AU. The model was shown to produce good results when comparing the magnetic field lines with the observed streamers (see \citet{Kuźma23}) and demonstrated a competitive run-time (typically 30 minutes to 2 hours) on high performance computing (HPC) setups, depending on the case complexity \citep{PerriLeitner2022}. 

Despite the demonstrated suitability of COCONUT for operational purposes, more work is needed in order to improve its physical accuracy and the reliability of its results. In particular, COCONUT results generally display very smooth density profiles showing little response to the electromagnetic structures such as streamers, despite the fact that these streamers are resolved accurately if one considers their magnetic field lines. Here, we show that this is largely caused by the pre-processing of the magnetic maps that we use for the mentioned simulations. Since we pre-process the magnetograms to enable numerical stability and fast convergence, by significantly reducing their resolution and magnetic field strength, we are also reducing the magnetic pressure within the computational domain. This causes an imbalance of forces and pressures yielding a relatively high plasma $\beta$. As a result, the force balance and thus the whole plasma dynamics is affected.

In this paper, we first briefly introduce the numerical setup of the solver in Section~\ref{sec:methods}, including the formulation, the grid, the boundary and initial conditions. In the same section, we also discuss the magnetic field pre-processing techniques and the expected magnetic field strength at our inner boundary, showing that our default setup uses magnetic maps that are too weak. In Section~\ref{sec:results}, we formulate an original technique to partly mitigate these effects, consisting in placing the inner boundary further away from the Sun while preserving the weaker magnetic field. We apply this technique on a dipole, a minimum and a maximum solar activity case, with validation against synthetic white-light imaging and tomography. We also evaluate the impact on the convergence behaviour, as a good performance and robustness remain essential for the operational utility of COCONUT, and discuss implications for the future use of the solver. We conclude with summarising our findings in Section~\ref{sec:conclusion}. 

\section{Methodology}
\label{sec:methods}

COCONUT is a plasma solver that was originally introduced by \cite{PerriLeitner2022}, with the aim of becoming an alternative, efficient MHD-based coronal model for the VSWMC. In this Section, we first present its formulation and the default numerical setup. Then we will focus on the prescription and pre-processing of the magnetic field on its inner boundary, highlighting the effects it has on the validity and resolution of the simulation results.

\subsection{Formulation}

COCONUT, a global coronal model based on the COOLFluiD Framework \citet{Lani2005, Kimpe2005, Lani2013}, solves for the non-dimensional ideal-MHD equations \citet{Yalim2011, Lani2014} with gravity using a fully-implicit unstructured second-order Finite Volume (FV) solver. The implicit nature of the steady-solver allows for CFL values much larger than 1, oftentimes up to tens to hundreds in an operational setting. The MHD formulation of the default COCONUT setup is given below:

\begin{equation}
    \frac{d\rho}{dt} + \mathbf{\nabla} \cdot (\rho \mathbf{V}) = 0,
\end{equation}

\begin{equation}
    \frac{d(\rho \mathbf{V})}{dt} + \mathbf{\nabla} \cdot \left( \rho \mathbf{V} \otimes \mathbf{V} + \mathbf{I} \left( P + \frac{1}{2}|\mathbf{B}|^2 \right) - \mathbf{B} \otimes \mathbf{B}  \right) = \rho \mathbf{g},
\end{equation}

\begin{equation}
    \frac{dE}{dt} + \mathbf{\nabla} \cdot \left( \left( E + P + \frac{1}{2}|\mathbf{B}|^2 \right)  \mathbf{V} - \mathbf{B} ( \mathbf{V} \cdot \mathbf{B})  \right) = \rho \mathbf{g} \cdot \mathbf{V},
\end{equation}

\begin{equation}
    \frac{d\mathbf{B}}{dt}  +  \mathbf{\nabla} \cdot \left( \mathbf{V} \otimes \mathbf{B} -  \mathbf{B}  \otimes \mathbf{V} + \mathbf{I} \phi \right) = \mathbf{0},
\label{eq:E10}
\end{equation}

\begin{equation}
    \frac{d\phi}{dt} + \mathbf{\nabla} \cdot \left( V^2_\text{ref} \mathbf{B} \right) = 0,
\end{equation}

where $\textbf{B}$ is the magnetic field, $\mathbf{V}$ the plasma velocity, $\rho$ the density, $P$ the scalar pressure, $\mathbf{g}$ the gravitational acceleration, $E$ the total internal energy, and $\phi$ serves for hyperbolic divergence cleaning. The reference values that are used to adimensionalise the equations are $B_\text{ref} = 2.2\cdot 10^{-4}\;$T,  $l_\text{ref} = 6.9551\cdot 10^8\;$m, and  $\rho_\text{ref} = 1.6\cdot 10^{-13}\;$kg/m$^3$. Note that in the equations above, radiation, conduction and heating are not included. The implementation of these terms into the solver formulation is currently ongoing. For the purpose of this study, we deem the baseline COCONUT formulation shown above to be sufficient since this baseline setup starts in the lower corona and resolves the magnetic structures well while also being very robust (see e.g. \cite{Kuźma23}).

\subsection{The grid}

The grid that was used in this study corresponds to the standard grid of the COCONUT solver, see \citet{Brchnelova2022a}. This grid is unstructured and based on a subdivided icosahedron spanning from $1.01\;R_\oplus$ to $25\;R_\oplus$, where the latter boundary was adjusted by \citet{Brchnelova2022b} to remove the outer boundary condition effects. For each of the cases shown below, a different grid resolution is used, with refinement increasing with the complexity of the flow field. The number of grid points for each case are shown in Table~\ref{tab:grid}.

   \begin{table}
      \caption[]{Grid resolution for the simulated cases.}
         \label{tab:grid}
     $$ 
         \begin{array}{p{0.5\linewidth}l}
            \hline
            \noalign{\smallskip}
            Case      &  \text{No. elements} \\
            \noalign{\smallskip}
            \hline
            \noalign{\smallskip}
            Dipole & 373.000     \\
           Minimum & 1.495.000             \\
            Maximum         & 1.946.000 \\
            \noalign{\smallskip}
            \hline
         \end{array}
     $$ 
   \end{table}

\subsection{Default initial and boundary conditions}

Since the solar wind at the outer boundary is super-fast, extrapolation from the last cell inside the domain is used at this boundary, and the inner boundary conditions fully determine the flow field. These are generally prescribed in terms of density, pressure, velocity and magnetic field. The magnetic field is set to represent the $B_r$ component of the magnetic map of the case that is to be studied. The density is generally prescribed to be 1.0 and the pressure to be 0.108, both in non-dimensional terms \citet{PerriLeitner2022}. For some cases, especially when maxima of solar activity are studied, the prescribed density might have to be locally or globally increased in order to ensure stability of the solver \citet{brchnelova2023assessing}.  The question of the magnitude and direction of the outflow velocity which is imposed on the inner surface was discussed in details in \citet{Brchnelova2022b}, as well as the initial conditions. In the following study, if the initial and boundary conditions deviate from what is indicated above, it will be clearly stated. 

\subsection{Magnetic map pre-processing}

 \begin{figure*}
   \centering
   \includegraphics[width=18cm]{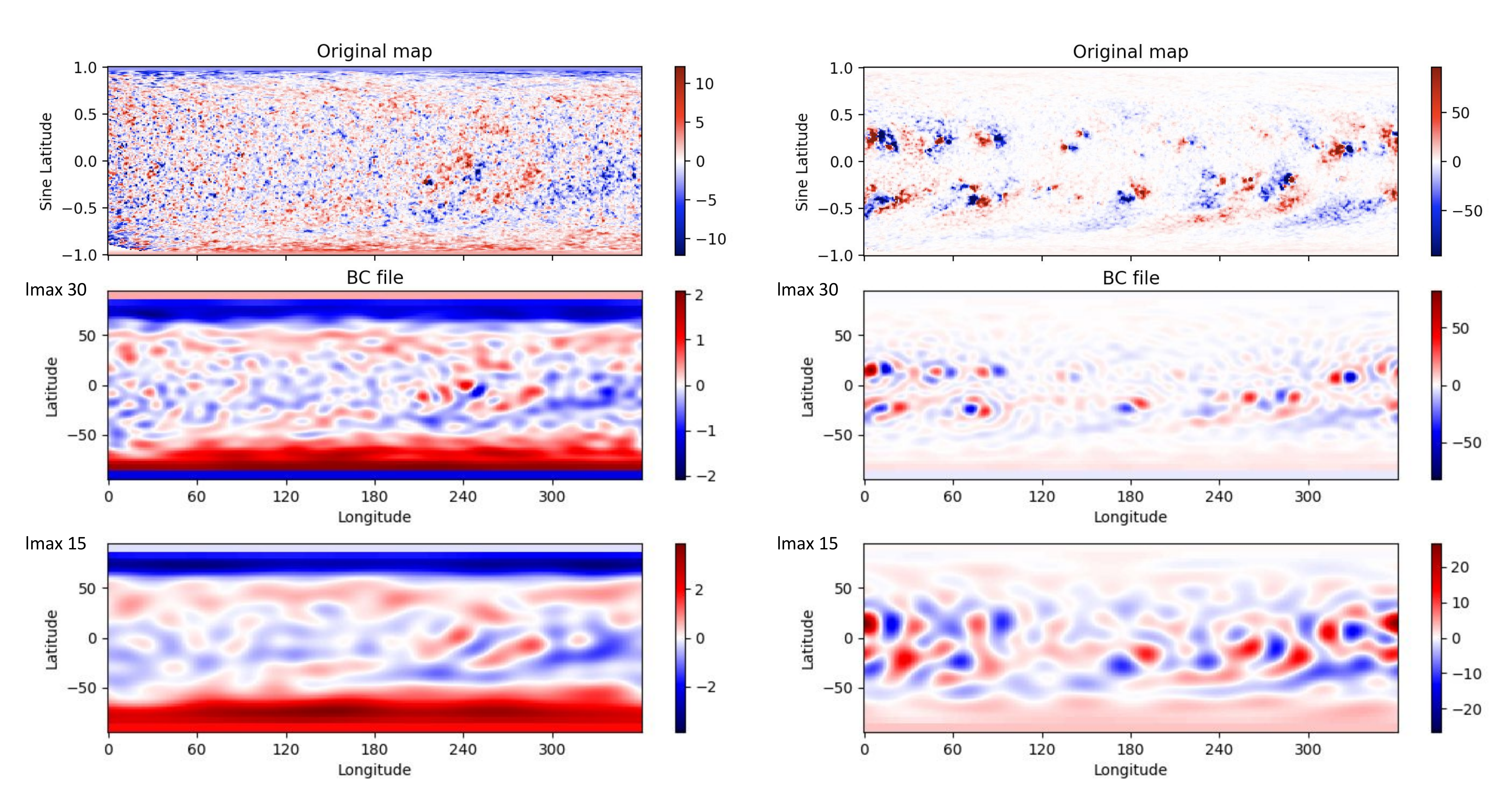}
   \caption{Demonstration of the pre-processing of the photospheric magnetic maps (top) through spherical harmonics processing, with $l_\text{max}$ of 30 (middle) and 15 (bottom). On the left, the magnetic map from the selected minimum of activity case ($1^{st}$ August 2008 eclipse) is shown. On the right, the magnetic maps correspond to the maximum of activity case ($13^{th}$ November 2012 eclipse). Units in Gauss.}
              \label{fig:blazejs}%
    \end{figure*}

    \begin{figure*}
   \centering
   \includegraphics[width=15.5cm]{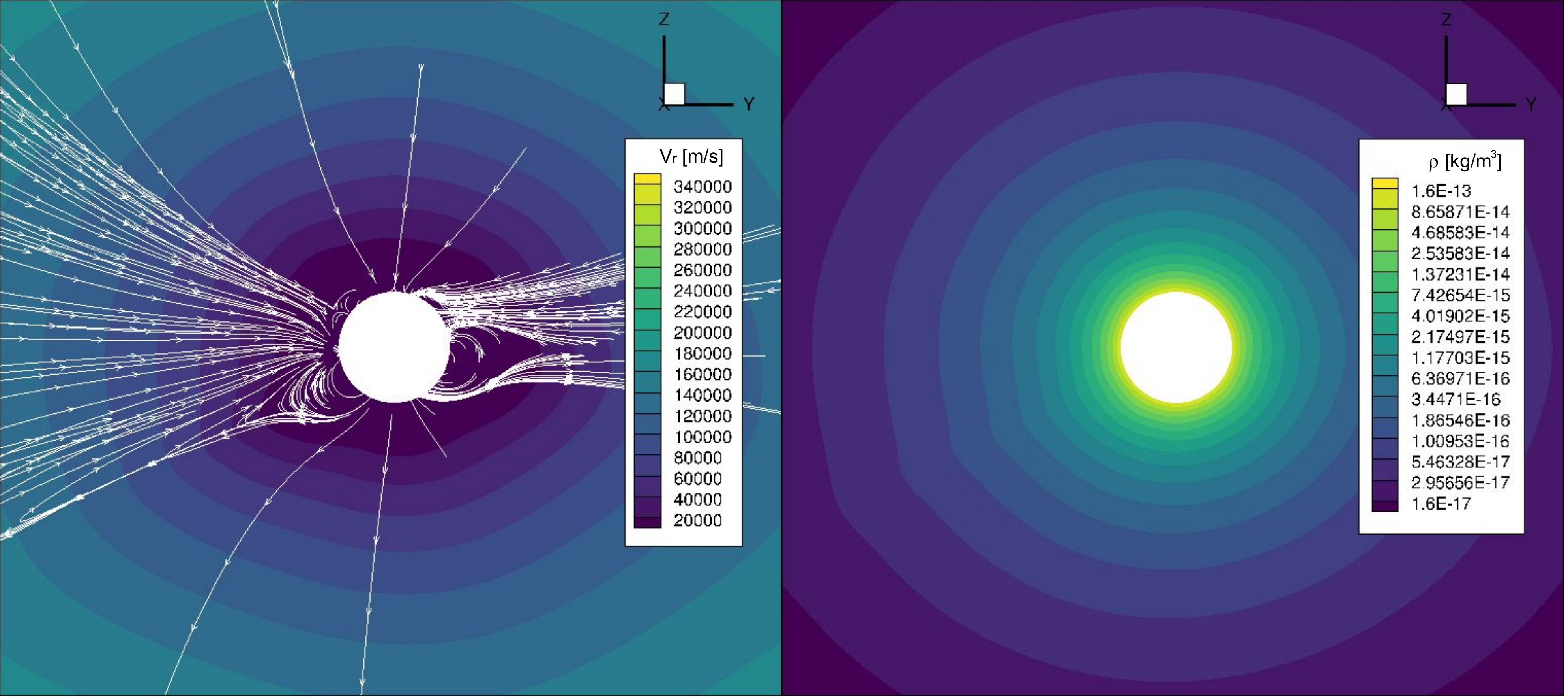}
   \caption{Resolved profiles of the 2008 minimum solar activity case. The left panel shows the magnetic field lines (in white) with the background contours indicating the radial velocity profile. The right panel shows the corresponding largely radial density profile.}
              \label{fig:minimumvrblinesdensity}%
    \end{figure*}

    \begin{figure*}
   \centering
   \includegraphics[width=15cm]{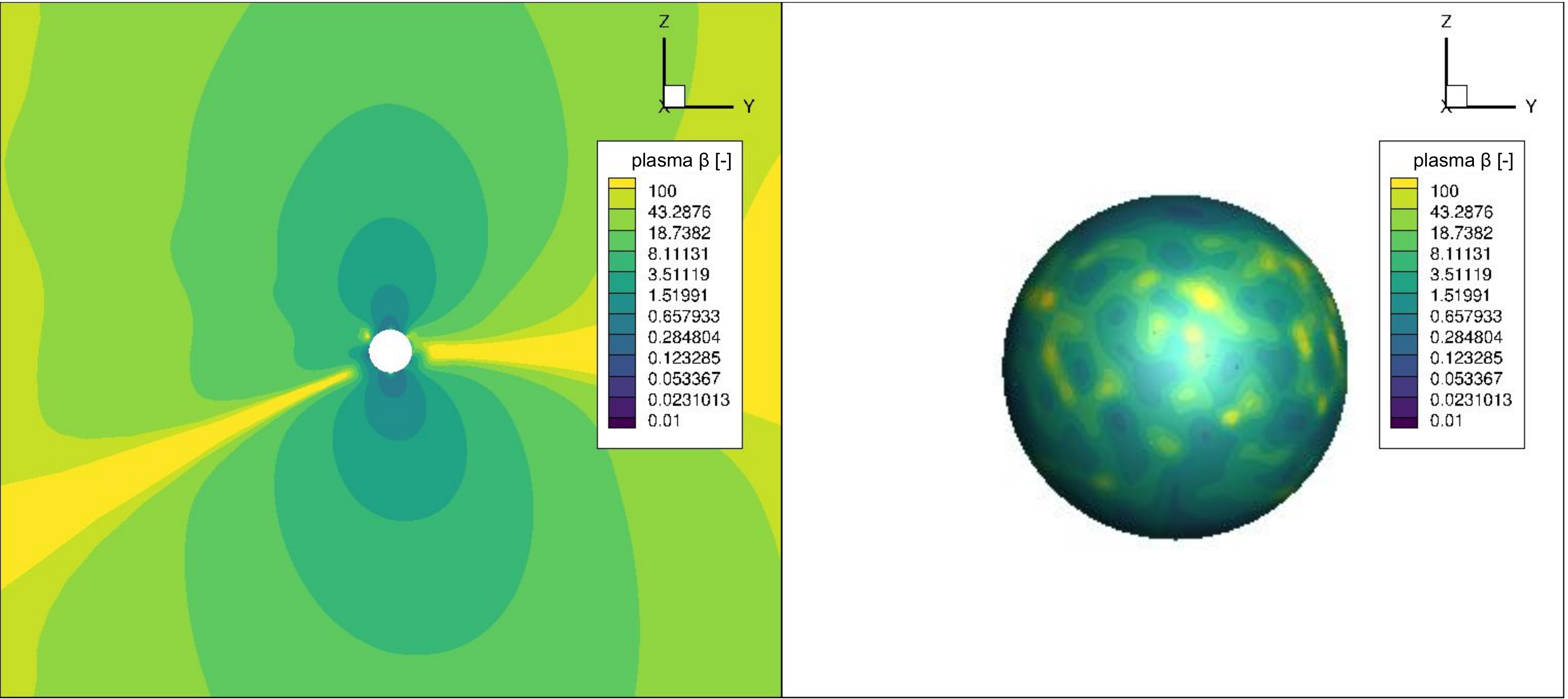}
   \caption{The plasma $\beta$ in the domain (left) and on the inner surface (right) for the 2008 minimum solar activity case.}
              \label{fig:minimumoriginalplasmabeta}%
    \end{figure*}

Photospheric magnetograms generally have very high resolution and very high magnetic field strengths. Due to the large gradients and the high local values of the magnetic field, unprocessed magnetograms are generally unpractical for global coronal simulations because they would easily affect their robustness and performance. In order to make them suitable for space weather forecasting runs, a smoothing is applied and this is frequently justified by the fact that the magnetic field is expected to be lower in the upper chromosphere and corona than in the photosphere (refer to e.g.\ the analysis of \citet{Joshi2017} for sunspots).

Therefore, our typical approach (see e.g.\ \cite{Kuźma23}) is to pre-process the photospheric magnetic maps through spherical harmonics, where the high spherical harmonics beyond a certain $l_\text{max}$ are filtered away. As a demonstration, refer to Figure~\ref{fig:blazejs}, where the separate magnetic maps correspond to the same original photospheric magnetogram shown on the top, but have different levels of  $l_\text{max}$. Both the magnetic field strength and the resolution diminish as $l_\text{max}$ decreases. In our setting, for rapid convergence, we generally consider  $l_\text{max}=15$ or 30, depending on whether the case corresponds to a minimum or maximum of solar activity, respectively. In the following, HMI magnetograms are used for our simulations unless stated otherwise \citep{Perri2022a}. 

Typically, this results in maximum magnetic field strength values in the range of 1-2\;G at the inner boundary for minima cases and 10-15\;G for maxima cases. Such values allow for a smooth run of the simulation, converging in times that are feasible for operational purposes (see \cite{PerriLeitner2022}). While the convergence performance is clearly one of the most important aspects of the solver as it is meant to become a part of a space weather forecasting tool chain, this limited strength of the magnetic field results in a decrease of the magnetic pressure and thus influences the level to which the plasma is capable of following the magnetic field lines. And while it is expected that the magnetic field and its gradients will diminish to a certain extent between the photosphere and the lower corona, this limit of $l_\text{max}$ was so far set mostly based on the operational performance of the solver, not on the physical evidence of the maximum magnetic field strength expected at the inner boundary. 

\subsection{Realistic {\bf B}-field strength in the lower corona and its effects}

Let us first look at the magnetic field that should be prescribed at the lower boundary. While we don't know exactly the magnetic field strength and evolution in the lower corona, there are several ways how this magnitude can be estimated from observations and simulations. A quite exhausting review was performed by \cite{AlissandrakisGary2021}, summarising the magnetic field strength evolution with height for active regions through different radio methods, such as polarisation reversal, metric bursts, or Faraday rotation, as summarised in Figure~14 of the referenced work. While there is quite a spread in the data due inclusion of the different techniques, it is clear that the estimated maximum magnetic field strength at $1.01\;R_\oplus$ should be much higher than the current 1-15\;G used in our simulations (with the zebra pattern and the Dulk-McLean relation showing a range of 50\;G to 200\;G). 

For a demonstration of this issue, we consider a case of a converged 2008 solar eclipse simulation. In Figure~\ref{fig:minimumvrblinesdensity}, we show the radial velocity profile with magnetic field lines (left) and the density profile (right), which clearly showcase the difficulty that the plasma has to follow the magnetic field lines. This insufficiency also presents itself in the evaluation of the plasma $\beta$. For the same 2008 case, we plotted the plasma beta in the flow field (left) and on the inner boundary (right) in Figure~\ref{fig:minimumoriginalplasmabeta}. If we take into account the estimations for plasma $\beta$, $\beta = n k T / (B^2 / 2\mu_0)$, from  Figure 3 of \cite{Gary2001}, we would be expecting the plasma $\beta$ at the location of $1.01\;R_\oplus$  to be below 0.01. While it is expected for this value to be very large along streamers where the magnetic field approaches very small values (i.e.\ the bright yellow regions), a sufficiently low plasma $\beta$ is clearly not achieved even in the rest of the domain or at the boundary. This means that the thermal pressure is too large compared to the magnetic pressure and explains why the resolved plasma dynamics lacks strong electromagnetic features.

\section{Results with a shifted boundary}
\label{sec:results}

From the discussion above, it is clear that the magnetic field which is currently prescribed for operational COCONUT runs is much smaller than what is realistic at that location in the low corona, at least when it comes to the active regions generally giving rise to strong features in the flow field. It is also clear that this assumption leads to deviations from what is physical in the simulation results. However, from our previous numerical experiments, even with increased limiting, prescribing a much higher magnetic field leads to deteriorated convergence and run-time performance. Therefore, an alternative solution should remedy this issue for operational usage. 

In order to improve the physicality of our simulations without prescribing higher magnetic fields however, we can invert the logic. Looking at the review paper of Alissandrakis and Gary (2021), Figure~14, the magnetic field strength of 1-10\;G that we assume would be expected roughly at $1\;R_\oplus$ height from the photosphere. Thus, we simply shift the inner boundary to this distance ($2\;R_\oplus$ in total). 

In order to be consistent, when shifting the inner boundary, we must adjust not only the magnetic field but also the other boundary conditions. To this end, we consider the profiles from the work of \cite{LemaireKatsiyannis2021}. According to Figure~1 of this paper, with the base equatorial density at the solar surface of around $10^{8.7}\;$cm$^{-3}$ ($8.4 \cdot 10^{-13}\;$kg/m$^3$), at $1\;R_\oplus$ height from the photosphere this number density drops to  $5.3 \cdot 10^{-15}\;$kg/m$^3$. The pressure is adjusted accordingly. 

{Note that there are topological consequences resulting from the fact that we are shifting the location at which we prescribe the magnetic field. These will be discussed in Subsection \ref{ss:validity}.}

\subsection{Dipolar case}

 \begin{figure*}
   \centering
   \includegraphics[width=15cm]{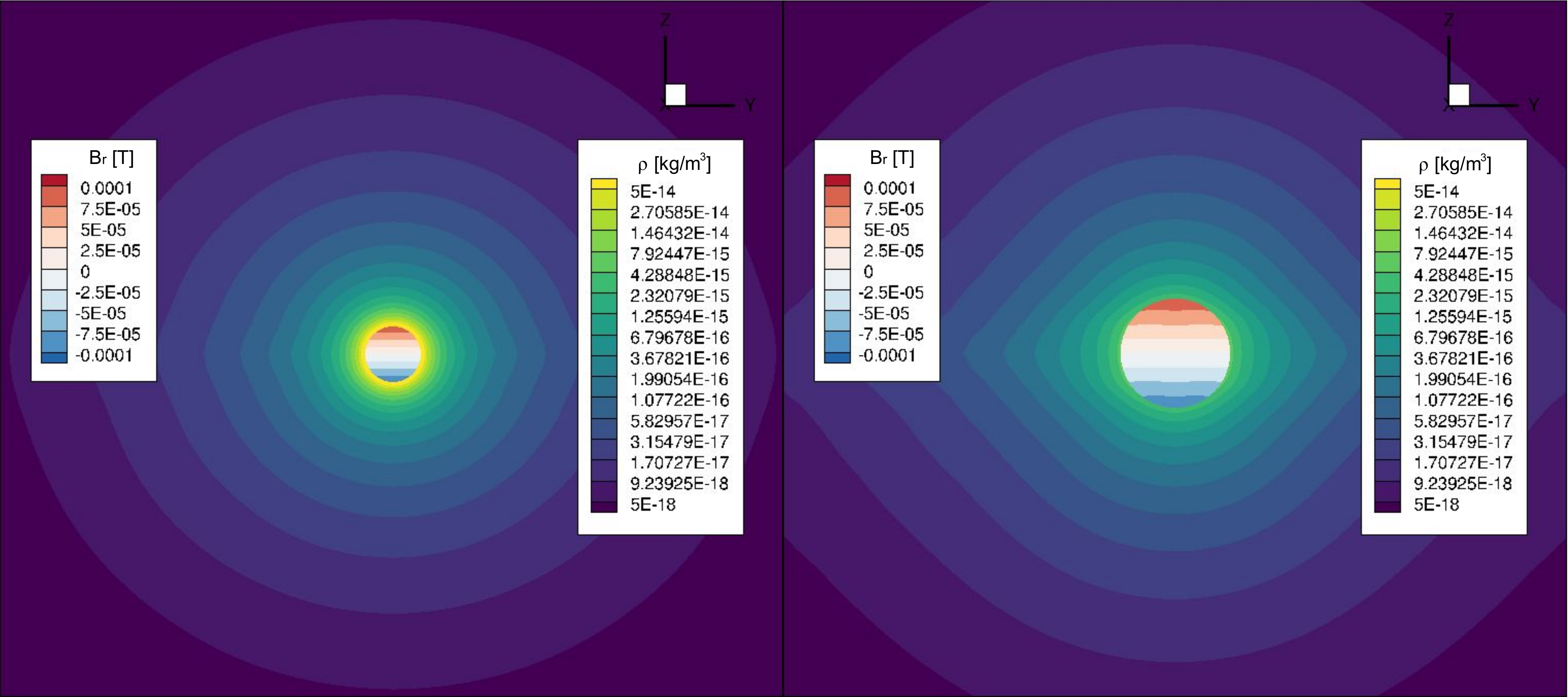}
   \caption{Density profiles of the simulated magnetic dipole with the original setup (left, boundary located at 1.01$\;R_\oplus$) and the new setup (right, boundary located at 2.0$\;R_\oplus$). The prescribed magnetic field is the same for both cases and shown on the inner boundary using the colour-bar on the left. The resolved density profile is indicated using the colour-bar on the right. }
              \label{fig:dipoleoriginalvsshifted}%
    \end{figure*}

 \begin{figure*}
   \centering
   \includegraphics[width=15cm]{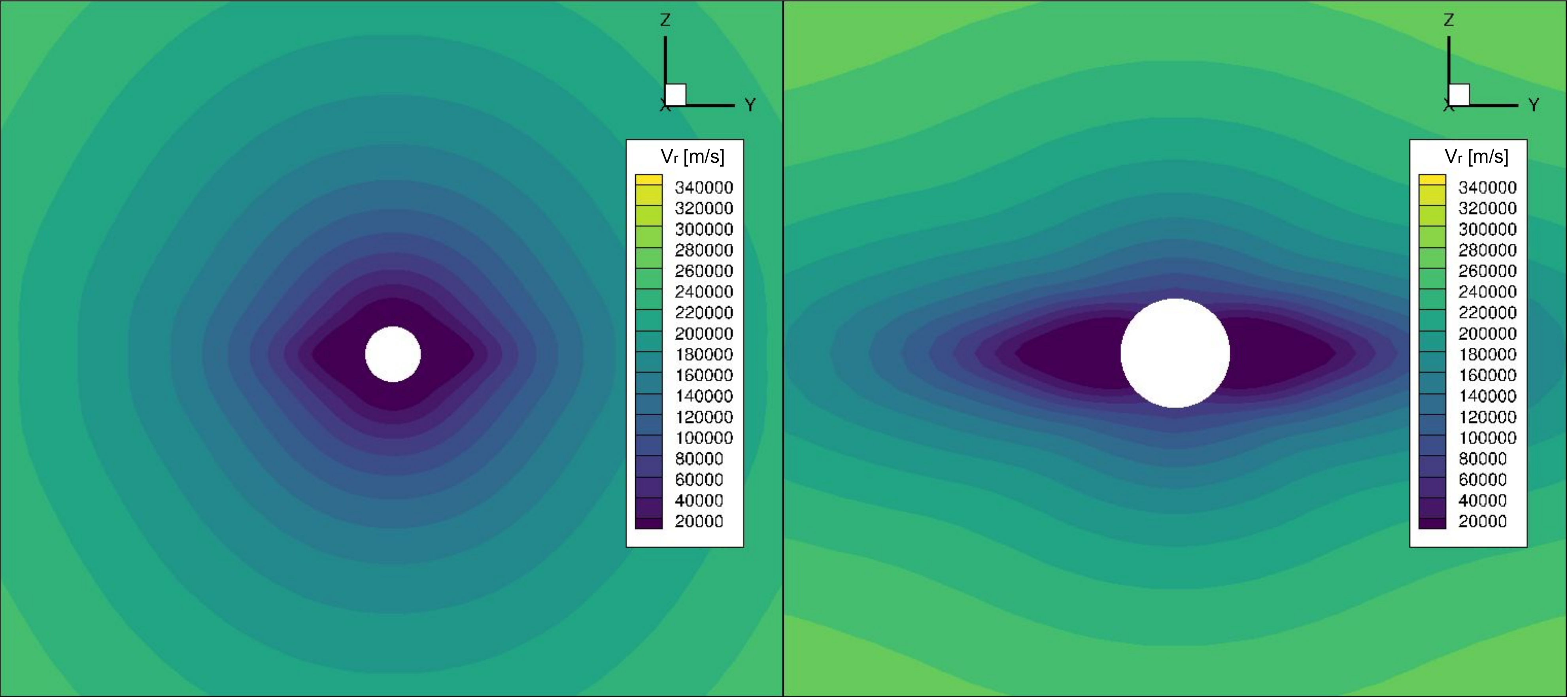}
   \caption{Radial velocity profiles of the simulated magnetic dipole with the original setup (left, boundary located at 1.01$\;R_\oplus$) and the new setup (right, boundary located at 2.0$\;R_\oplus$). }
              \label{fig:dipolevr}%
    \end{figure*}

First, we apply this logic to the case of a dipole to see how this modification affects the density field in a simple case. In Figure~\ref{fig:dipoleoriginalvsshifted}, the density profiles of the original approach (on the left) and the shifted-boundary approach (on the right) are demonstrated. The surface magnetogram sphere in the middle in the red and blue colour-map demonstrates that while the inner sphere on the left is 1.01$\;R_\oplus$ and on the right 2$\;R_\oplus$, the magnetic field that is prescribed on it is the same, following the logic outlined above. The Figure shows that the new approach leads to more pronounced density features around the equator as would be expected from a case in which the electromagnetic forces and the magnetic pressure are more dominant compared to the gravity and the thermal pressure. 

A similar observation can be made from plotting the radial velocity, in Figure~\ref{fig:dipolevr}, where again the original case is on the left and the shifted-boundary case on the right. Despite the fact that the range for the velocity magnitude in the domain remains the same, the equatorial streamers are much better shaped in the shifted-boundary case. 

In order to verify whether these observed changes are significant enough to improve also the simulations based on real magnetic maps, this approach is also tested for a solar minimum and a solar maximum case.

\subsection{Minimum and maximum activity cases}

 \begin{figure*}
   \centering
   \includegraphics[width=15cm]{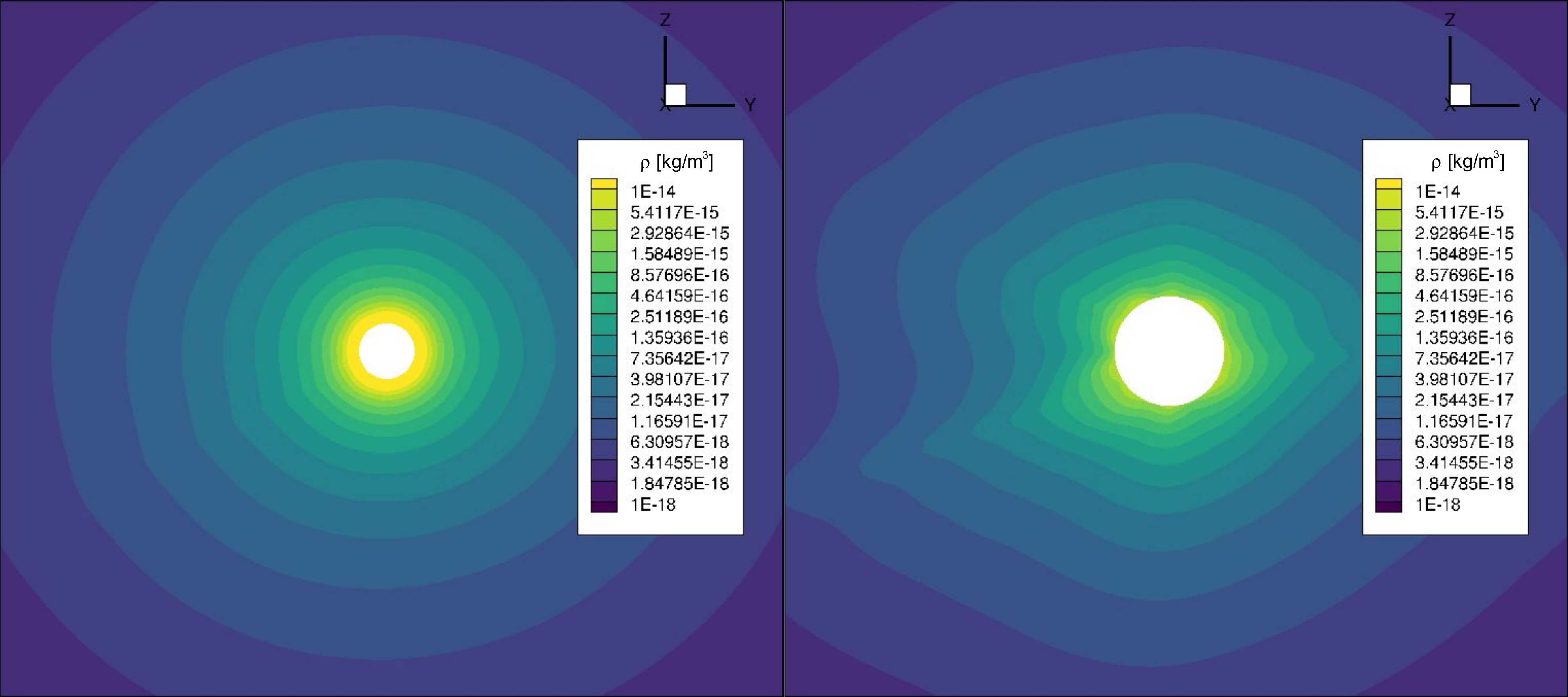}
   \caption{Density profiles of the 2008 minimum solar activity case with the original setup (left, boundary located at 1.01$\;R_\oplus$) and the new setup (right, boundary located at 2.0$\;R_\oplus$).}
              \label{fig:minimumoriginalvsshifted}%
    \end{figure*}

    \begin{figure*}
   \centering
   \includegraphics[width=15cm]{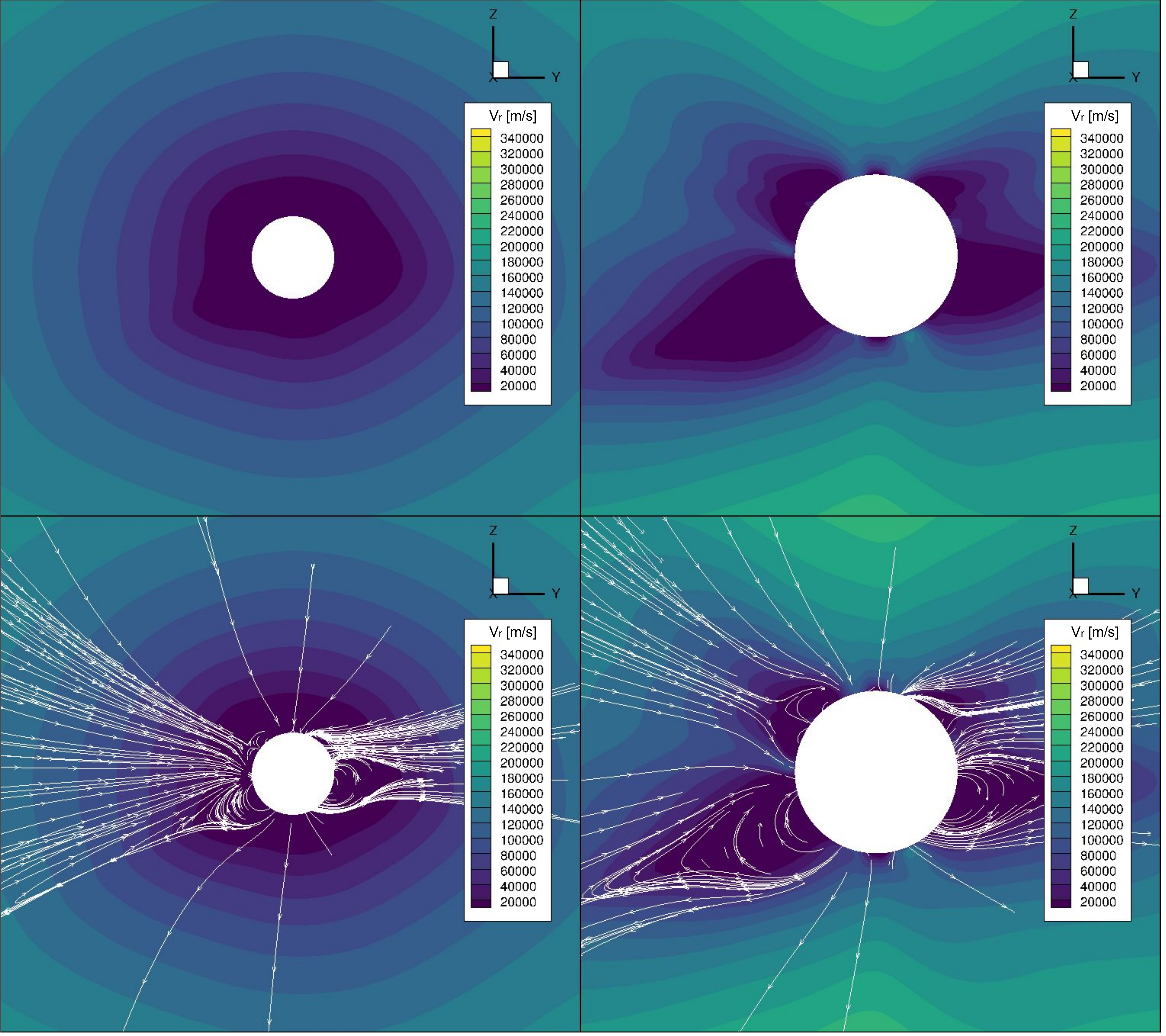}
   \caption{Profiles of the 2008 minimum solar activity case with the original setup (left, boundary located at 1.01$\;R_\oplus$) and the new setup (right, boundary located at 2.0$\;R_\oplus$). The top two panels show the radial velocity profiles alone, while the bottom panels also overlay the magnetic field lines (in white) to help ascertain how well the plasma flow follows them.}
              \label{fig:minimumvr}%
    \end{figure*}
    
In applying this new approach to real cases, we start with the 2008 eclipse case which was previously introduced. We repeat the steps described for the dipole, shifting the inner boundary to 2$\;R_\oplus$ and adjusting the boundary density and pressure accordingly while keeping the same prescribed magnetic map. The results for this and the original approach are shown in Figure~\ref{fig:minimumoriginalvsshifted} which demonstrates that the density profile in the case of the shifted-boundary approach (right) is much better developed compared to the original (left). 

Comparison of radial velocity profile, in Figure~\ref{fig:minimumvr}, for the original simulation (left) and the shifted-boundary simulation (right) portrays a much higher level of detail for the latter. Based on the investigation of the shape of the magnetic field lines in the bottom panels of Figure~\ref{fig:minimumvr}, which were already validated in \cite{Kuźma23}, it can be concluded that the added detail is physical. Considering that both simulations are resolved using the same grid refinement and the same scheme, this added level of detail is fairly significant and can help us reduce the computational cost of predictions by allowing us to potentially run simulations on even coarser grids. 

\begin{figure*}
   \centering
   \includegraphics[width=15cm]{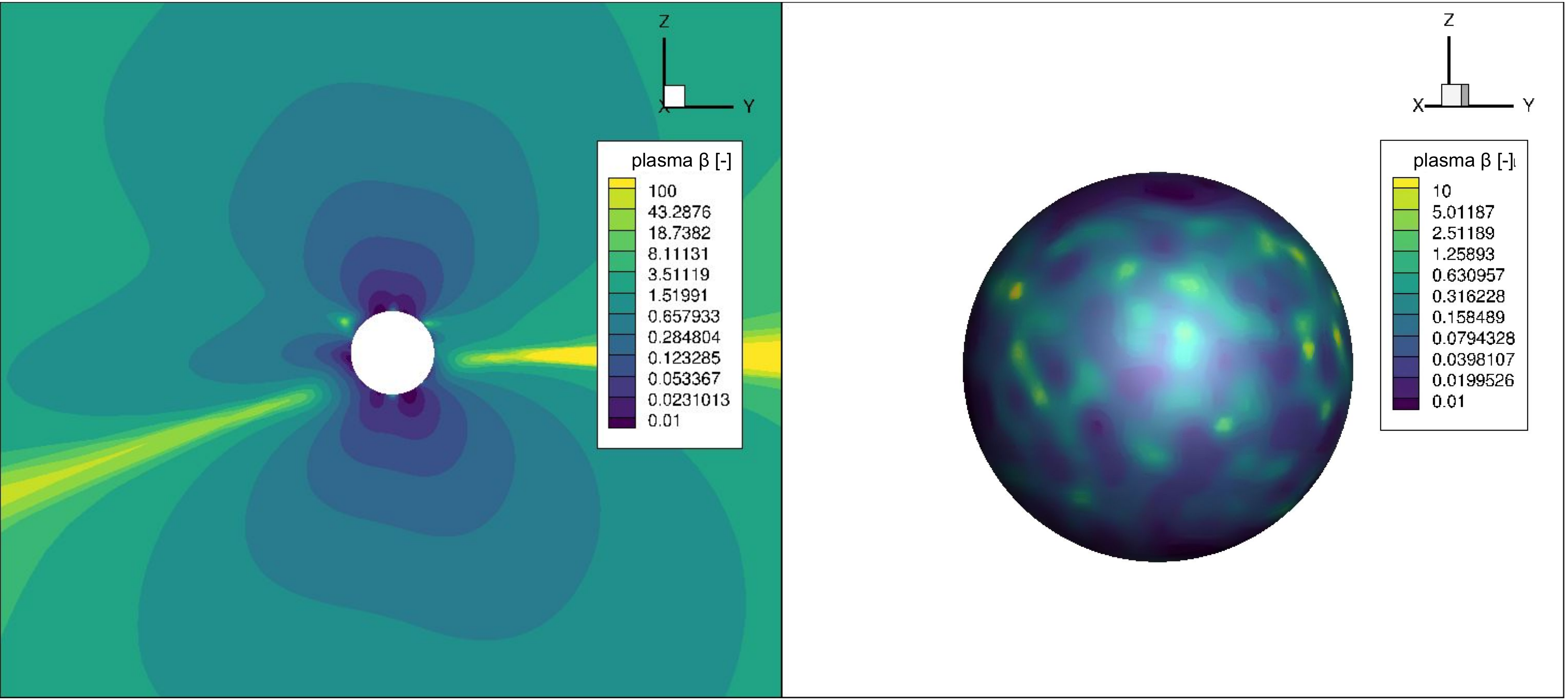}
   \caption{The plasma $\beta$ in the domain (left) and on the prescribed surface (right) for the 2008 minimum solar activity case, as resolved using the new approach of shifting the boundary to 2$\;R_\oplus$.}
    \label{fig:minimumshiftedplasmabeta}%
    \end{figure*}

This enhancement can also be understood as an improvement in matching the correct plasma $\beta$ both at the inner boundary and in the flow field. The plasma $\beta$ for the shifted-boundary simulation is shown in Figure~\ref{fig:minimumshiftedplasmabeta}. {If we investigate the plasma $\beta$ in Figure 3 from \cite{Gary2001}, over the active regions at the distance of 2$\;R_\oplus$, we would expect plasma $\beta$ values in the range of 0.3 to around 15 (nondimensional)}. This is similar to the values obtained at the inner boundary with this approach. In the rest of the flow field, we would expect the value to keep increasing up to more than 0.1 at the outer boundary, which again is satisfied. 

From the analysis above and revisiting the ideal-MHD formulation, it is possible to recognise the following three ways in which shifting the inner boundary helps the solver to arrive to a more physical and a better resolved flow field:

\begin{itemize}
    \item by assuming a smaller boundary density at 2$\;R_\oplus$, the thermal pressure and the gravity are automatically reduced while the magnetic pressure is kept the same, 
    \item by increasing the distance from the centre of the Sun, the gravity force is reduced even further, and
    \item as the boundary on 2$\;R_\oplus$, the higher plasma $\beta$ is more physical.
\end{itemize}

To further showcase that the more detailed density profile leads to more physical results, synthetic WLI were generated from these simulations for both cases. For synthetic WLI, we followed the spherically symmetric inversion method of \cite{Billings1966} in order to compute, for each point, the polarised brightness is {integrated along the line-of-sight (LOS) as}:

\begin{equation}
    pB \propto \int_{\text{LOS}} \Big( (1 - u) A(r) + u B(r) \Big) \frac{\rho^2}{r^2} N(s) ds,
\end{equation}

\noindent where,

\begin{equation}
    A(r) = \cos \Omega \sin^2 \Omega, 
\end{equation}

\begin{equation}
    B(r) = - \frac{1}{8} \Big[ 1 - 3 \sin^2 \Omega - \cos^2 \Omega \Big( \frac{1 + 3 \sin^2 \Omega}{\sin \Omega} \Big) \ln \Big( \frac{1 + \sin \Omega}{\cos \Omega} \Big)  \Big],
\end{equation}

\noindent and $\sin \Omega = \frac{R_\oplus}{r}$ with $r^2 = s^2 + \rho^2$.

{In the expressions above, $s$ represents the path distance over which the integration is carried out, $\rho$ the perpendicular distance to the solar disk and} $u$ is the limb darkening coefficient, which was set to 0.5 in the present study. 

The box over which the ray tracing was performed was set to 3.5$\;R_\oplus$ x 3.5$\;R_\oplus$ x 3.5$\;R_\oplus$ with 250 x 250 rays, originating from uniformly distributed points. Convergence study on the size of the domain and number of rays revealed that these values were sufficient to accurately capture the brightness features (or lack thereof). 

    \begin{figure*}
   \centering
   \includegraphics[width=18cm]{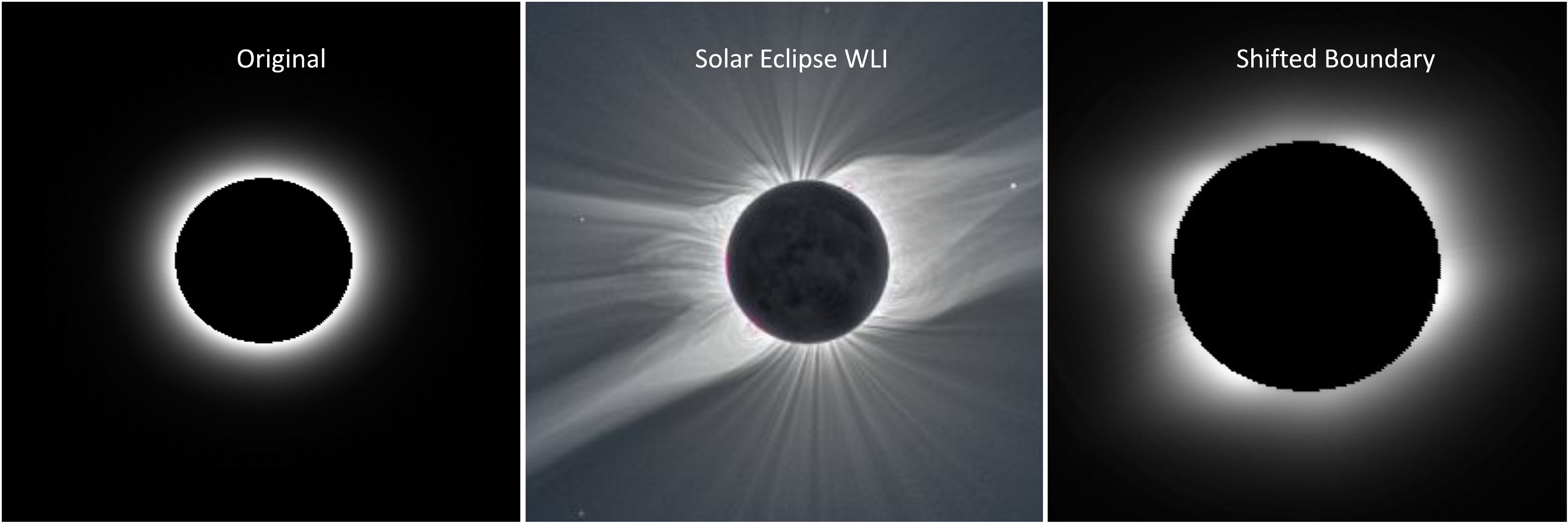}
   \caption{Synthetic WLI for the 2008 minimum solar activity case using the density profiles from the original approach (left, boundary located at 1.01$\;R_\oplus$), the new setup (right, boundary located at 2.0$\;R_\oplus$) and the corresponding eclipse observation  (© 2008 Alson Wong) in the middle.}
              \label{fig:minimumwlishifted}%
    \end{figure*}

 \begin{figure*}[h!]
   \centering
   \includegraphics[width=19cm]{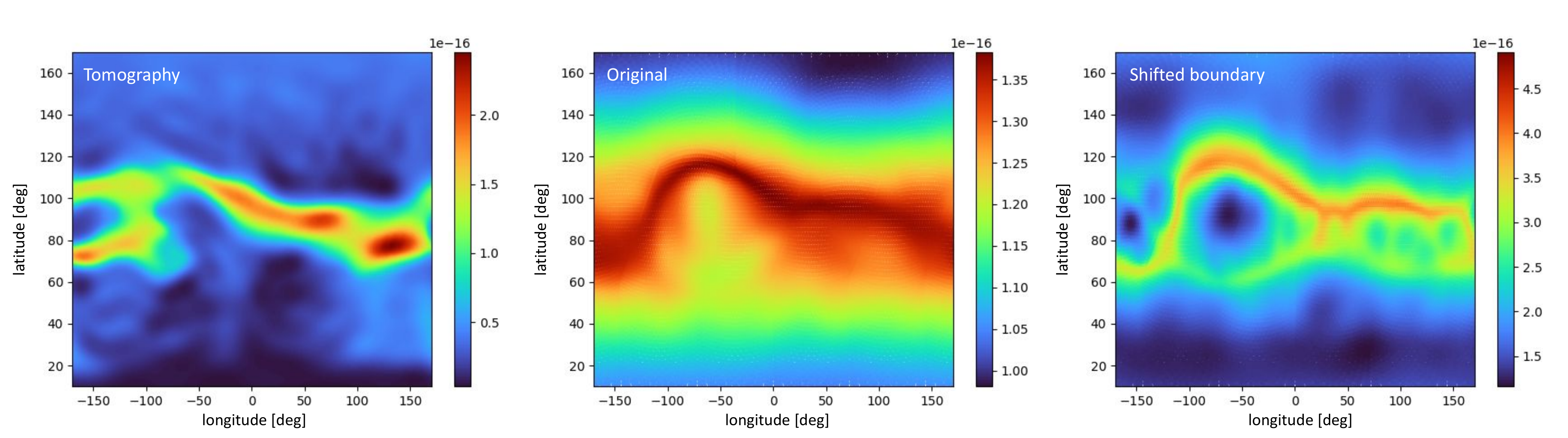}
   \caption{Comparison of tomography measurement of the 2008 minimum solar activity case (left) from Morgan et al. with synthetic tomography using the density profiles from the original approach (middle, boundary located at 1.01$\;R_\oplus$) and the new setup (right, boundary located at 2.0$\;R_\oplus$).}
              \label{fig:tomography}%
    \end{figure*}

 \begin{figure*}
   \centering
   \includegraphics[width=15cm]{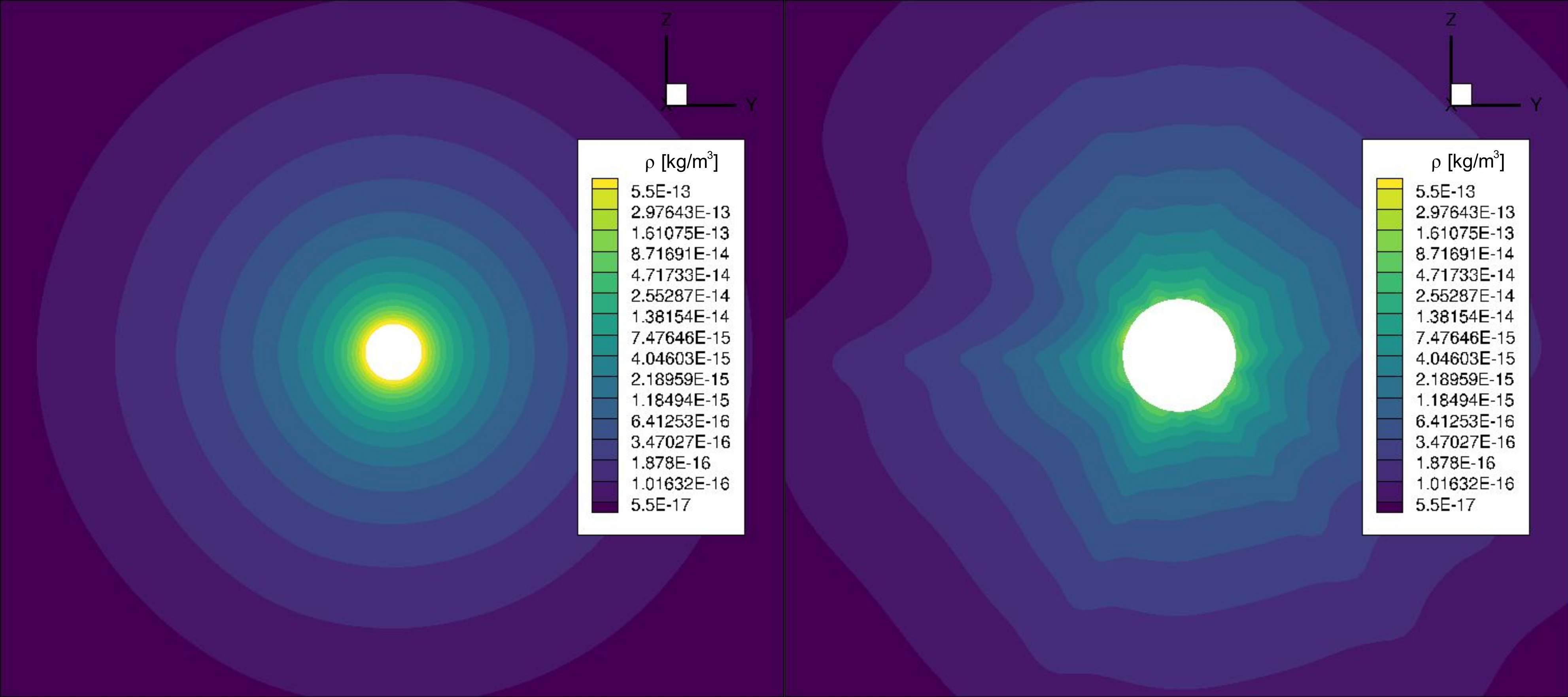}
   \caption{Density profiles of the 2012 maximum solar activity case with the original setup (left, boundary located at 1.01$\;R_\oplus$) and the new setup (right, boundary located at 2.0$\;R_\oplus$).}
              \label{fig:maximumoriginalvsshifted}%
    \end{figure*}

    \begin{figure*}
   \centering
   \includegraphics[width=18cm]{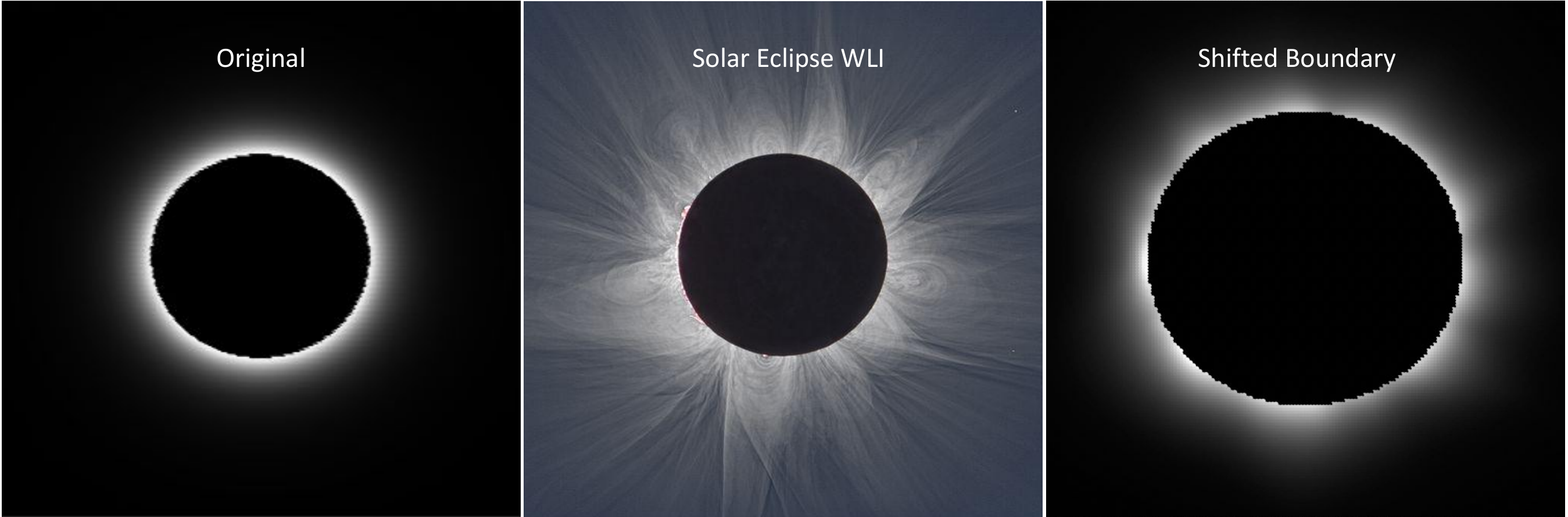}
   \caption{Synthetic WLI for the 2012 maximum solar activity case using the density profiles from the original approach (left, boundary located at 1.01$\;R_\oplus$), the new setup (right, boundary located at 2.0$\;R_\oplus$) and the corresponding eclipse observation  (© 2012 Constantine Emmanouilidi, Miloslav Druckmuller) in the middle.}
              \label{fig:WLImaxshifted}%
    \end{figure*}
    
The results of WLI are shown in Figure~\ref{fig:minimumwlishifted}, where the original approach is shown on the left, the new approach on the right, {and the observation in the middle. It should be remarked that it is not the intention of the authors to have the reader thoroughly compare the exact contrast, location and geometry of the visible synthetic WLI features with the WLI observations since it is known that inclusion of additional physical terms, which are currently neglected, will further affect the positioning and shape of these streamers (and hence such a detailed comparison with the real physics would be unfair for both approaches). The contrast and definition in the shown WLI observation is also achieved through sophisticated post-processing, currently not yet implemented into our synthetic WLI image generation. However, the observation is included nevertheless to get an indication of where one should expect intensity enhancements for this specific solar activity level and magnetic field configuration.} In the original case (left), no features are visible, which is why the comparison with the eclipse pictures through WLI was previously impossible for COCONUT simulation results. On the right, for the new approach, enhancements of intensity are seen in the locations where the streamers are located shown in Figure~\ref{fig:minimumvr}, in similar locations as the enhancements seen in the observation in the middle. As stated above, it is expected that the inclusion of more physics into our model, such as coronal heating approximation, wave turbulence terms, conduction and radiation and/or various WLI filters and post-processing will enhance the features even further.  

Besides WLI, another mean of validating (at least qualitatively) the shape of the resolved density profiles for cases of minimum solar activity is tomography. From tomography, the profiles of electron density can be derived at various distances from the Sun, generally between 4 to 8 $\;R_\oplus$. We used the coronagraph image data at 4.4 $\;R_\oplus$ of \citet{Morgan15} after having processed it through a spherical harmonic-based regularised inversion method \citep{Morgan19} and further refinement steps \citep{Morgan20}.

The tomography for the date of the minimum of activity case is shown in Figure~\ref{fig:tomography} on the left. In the middle of the same Figure, the synthetic tomographic image was created using the original setup with a weak magnetic field applied on 1.01$\;R_\oplus$. On the right, the synthetic tomographic image was generated for the new approach, with the boundary shifted to 2$\;R_\oplus$. It is clear that the new approach produces a greater level of detail which is also generally in agreement with the actual observation, despite the fact that the model is still only polytropic. 

Next, we apply this approach also to a case of maximum of solar activity, to demonstrate that this methodology can be also used for more complex cases. Here, we take the magnetic map of the previously mentioned 2012 eclipse and repeat the same processes as for the minimum case above.

The resulting density profiles for the original approach (left) and the shifted-boundary simulation (right) are shown in Figure~\ref{fig:maximumoriginalvsshifted}, again showing a much better development of the density profile, following the magnetic lines in the domain.  

Similarly, the synthetic WLI were generated for comparison, through the same method as outlined above. Synthetic WLI and {the observation of the corresponding solar eclipse are} shown in Figure~\ref{fig:WLImaxshifted}, demonstrating much more pronounced brightness features for the new approach compared to the original, located in the regions of streamers and density enhancements. 

While it is possible to somewhat improve the features of the maxima cases by using higher resolution magnetograms using the original setup with adjustments in the divergence cleaning method, these oftentimes take much longer to converge, lead to nonphysically high-velocity patches or generate unstable features which prevent the simulation from meeting the residual criteria for convergence. The approach with the shifted boundary does not face these problems.

\subsection{The validity of the prescribed {\bf B}-field strength at 2$\;R_\oplus$}
\label{ss:validity}

\begin{figure}
  \resizebox{\hsize}{!}{\includegraphics[width=4cm]{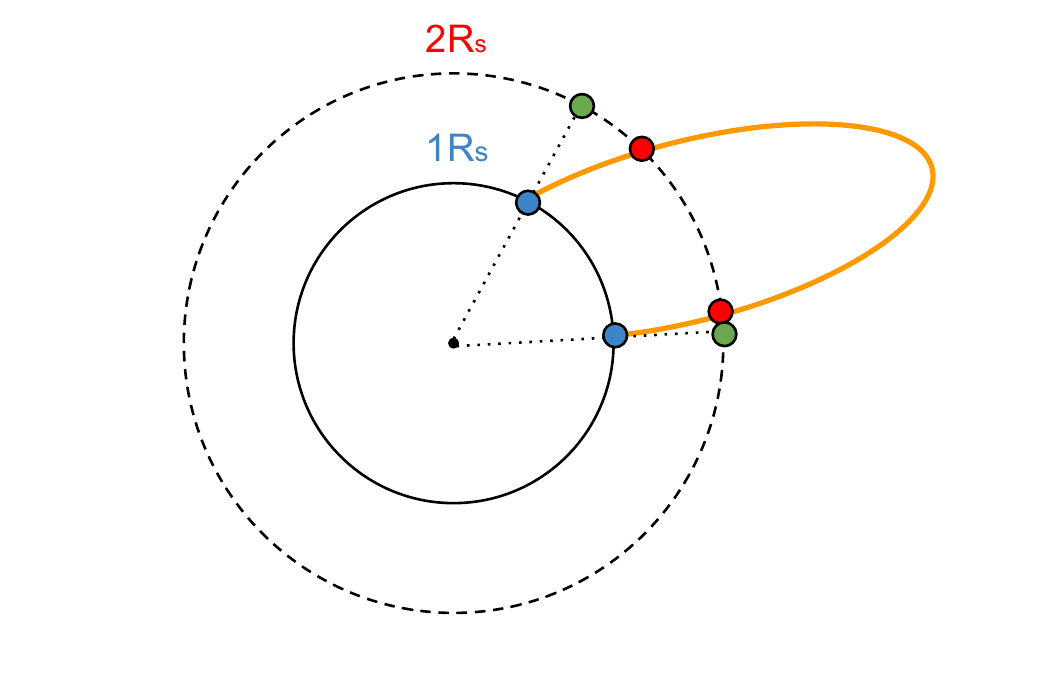}}
  \caption{The effects of prescribing the magnetic field at 2$\;R_\oplus$ (dashed black circle) instead of 1$\;R_\oplus$ (solid black circle) on the loop size and geometry (orange solid line).}
  \label{fig:EffectsOfMovingBoundary}
\end{figure}

While prescribing a filtered photospheric magnetic field at 2$\;R_\oplus$ might be more realistic in terms of the magnetic field strength and gradients present in it, the fact that the loop footpoints are shifted {by one solar radius} might change the magnetic field topology, see Figure~\ref{fig:EffectsOfMovingBoundary}. In the Figure, the original, "true" footpoints of a given loop (solid orange arc) are located on the 1$\;R_\oplus$ boundary (solid circle) in blue points. As this loop moves to the region of 2$\;R_\oplus$ (dashed circle), the imagined footpoints would move to the red points. However, by simply prescribing the same magnetic field at 2$\;R_\oplus$, we artificially move the footpoints to the green points through a simple linear upscaling (dotted black lines), thus, making the structure larger. 

In addition, some of the original 1$\;R_\oplus$ loops might be small enough that they do not even extend to the distance of 2$\;R_\oplus$. Obviously, by using the same magnetogram as for 1$\;R_\oplus$ at 2$\;R_\oplus$, these small loops will still form at this distance. 

 \begin{figure*}
   \centering
   \includegraphics[width=18cm]{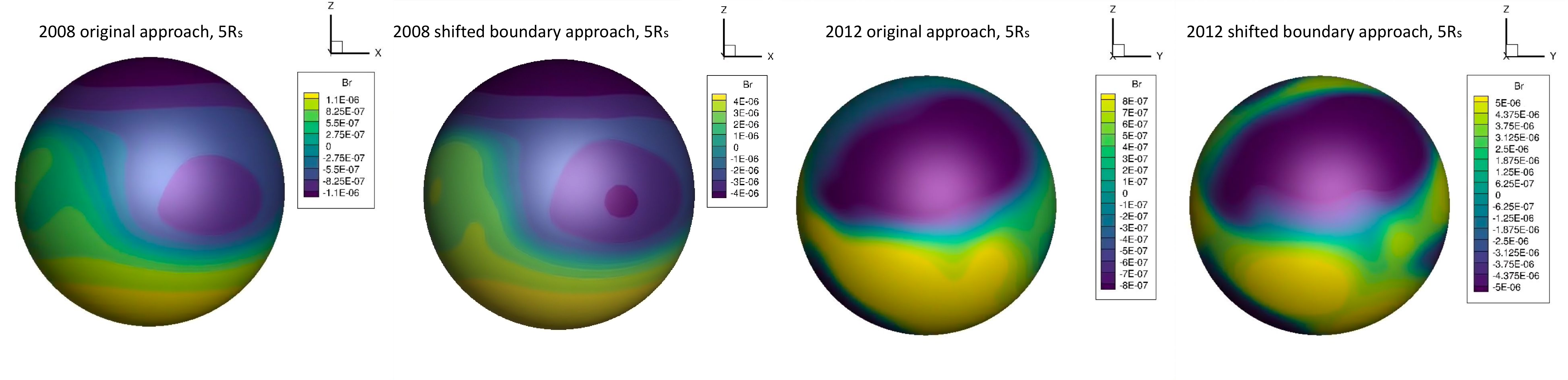}
   \caption{Comparison of the resulting $B_r$ structures of the different approach at 2$\;R_\oplus$ for 2008 (left two) and for 2012 (right two).}
              \label{fig:BrAt5RsBoth}%
    \end{figure*}

The implication of all of this is that:

\begin{itemize}
    \item the modelled loops might end up being larger with a different geometry, depending on their position and shape, and
    \item "spurious" loops might exist close to the 2$\;R_\oplus$ surface, even though their extent would otherwise be limited a distance of much less than 2$\;R_\oplus$. 
\end{itemize}

Despite this, however, as shown in the previous section, we still get much better profiles when it comes to e.g.\ synthetic tomography, indicating sharper and more accurate density features further away from the Sun. There are several reasons why this is the case, namely:

\begin{itemize}
    \item more pronounced ("upscaled") electromagnetic structures and lower dissipation result in that the numerical reconnection in the streamers and current sheets occurs much further away from the star, meaning that the resulting plasma features are much better pronounced as they extent into the rest of the domain,
    \item despite the fact that the geometry and size of the loop might be affected due to the above-mentioned reasons close to the inner surface, the general position of the apex of the loop is still sufficiently accurate to create density enhancements in the right places, and
    \item what matters for space weather forecasting are the conditions at 0.1\;AU, not at 2$\;R_\oplus$, thus the effect of the inaccuracies close to the surface is diminished.
\end{itemize}

{The above points are demonstrated in Figure~\ref{fig:BrAt5RsBoth}, where the $B_r$ features at the 5$\;R_\oplus$ surface are shown. It is clear that despite the different treatment of the inner boundary and different resulting velocity and density profiles in the domain, the magnetic field structure further away from the Sun looks very similar thanks to premature reconnection and the fact that the loop apexes in the new approach are still positioned at the right places. The only difference comes in terms of the magnitude of the B-field, which can however be expected since in the new approach, the same magnetic field strength is prescribed at a larger distance into the domain.}

While this definitely makes it difficult to use the results of these simulations for detailed modelling of the behaviour of the coronal loops and other structures close to the star, thanks to the reasons mentioned above, we believe that this "upscaling" of the electromagnetic features might still be a suitable technique for improving space weather forecasting.

 \begin{figure*}
   \centering
   \includegraphics[width=15cm]{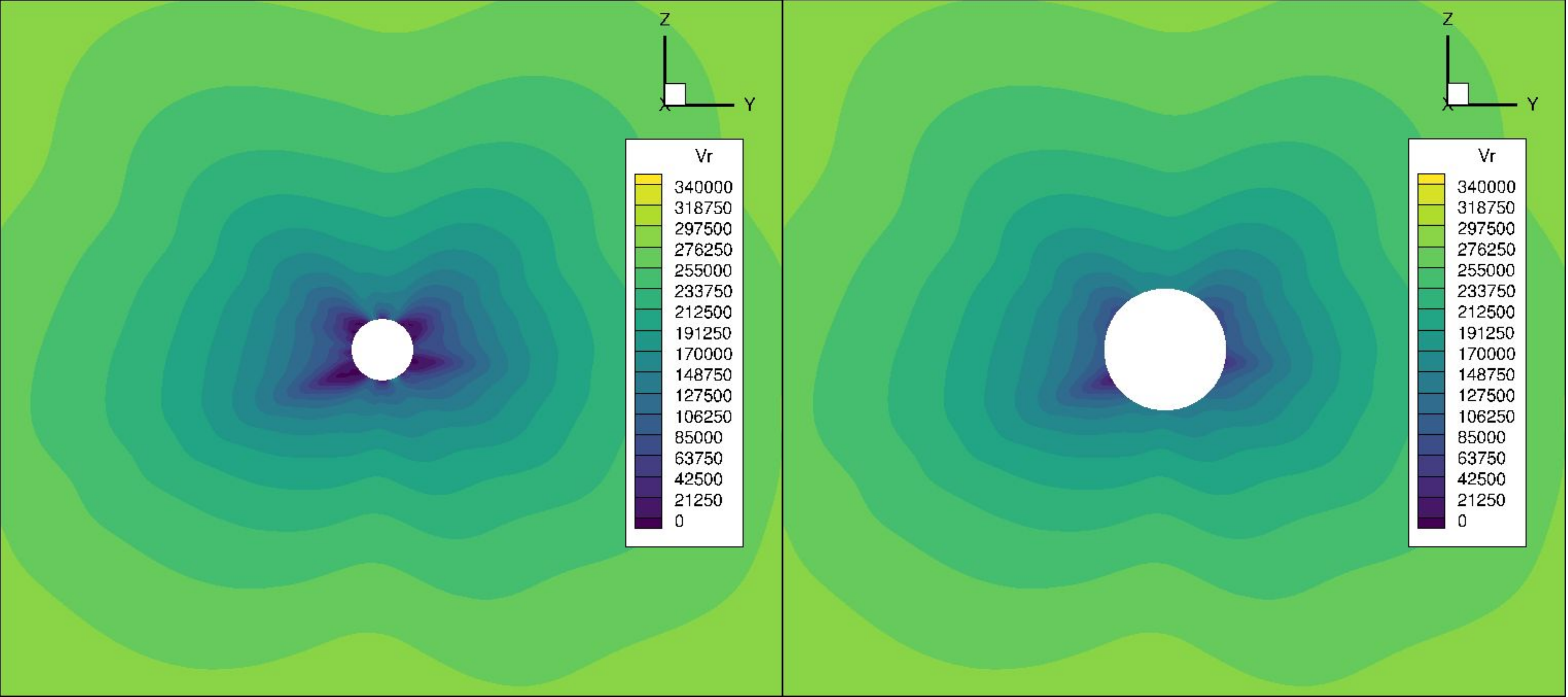}
   \caption{Comparison of the two interpretations of the adopted approach. Both subfigures show the radial velocity profile. The left-hand side plot shows the general result of the simulation, where the inner boundary can represent either a 2$\;R_\oplus$ or a 1$\;R_\oplus$ surface. On the right, the plot shows where the 2$\;R_\oplus$ surface would be located if the left-hand side plot's boundary would be at 1$\;R_\oplus$.}
              \label{fig:1Rsvs2Rs}%
    \end{figure*}
    
An alternative interpretation of the results that solves this problem exploits the non-dimensional nature of the MHD equations. Indeed, the results of the shifted-boundary case with the larger grid can be interpreted as the results computed on the original 1$\;R_\oplus$ domain with a decreased boundary density, pressure and gravity. In that case, the only parameter that changes between the two simulations in practice is the reference length, which is the reason why the gravitational force must be down-scaled. In both cases, as also numerically verified, the results look exactly the same and for the 2008 minimum case, the corresponding $V_r$ field is shown in the left-hand side plot of Figure \ref{fig:1Rsvs2Rs}. However, with the 2$\;R_\oplus$-boundary approach, it is implied that the features shown in the left-hand side plot of Figure \ref{fig:1Rsvs2Rs} correspond to the features that would originate form the 2$\;R_\oplus$ boundary. In contrast, if we assume that the simulation is solved on the original 1$\;R_\oplus$ grid but with an artificially decreased thermal pressure and gravity, the features shown in the left-hand side plot of Figure \ref{fig:1Rsvs2Rs} would be assumed to originate from a 1$\;R_\oplus$ boundary, and the features around the 2$\;R_\oplus$ surface in the same domain would look like the features depicted in the right-hand side plot of Figure \ref{fig:1Rsvs2Rs}. The fact that the features are much larger around the 2$\;R_\oplus$ surface on the left (according to the original interpretation) compared to the 2$\;R_\oplus$ surface (according to the alternative interpretation) on right further demonstrates the "upscaling" of discussed above. 

The question then becomes which one of these interpretations is more accurate and therefore, whether the physical scale of the domain should be doubled in our simulations or not. On one hand, the latter interpretation does not face the upscaling problem and resolves sharp features with an appropriate plasma $\beta$ close to the surface of the Sun. On the other hand, the former interpretation assumes a correct thermal pressure near the surface and a correct gravitational force. Considering that the scaled-down gravitational forces influence density profiles over the entire domain and hence lead to incorrect density gradients especially further away from the Sun where gravity dominates (and where the solution would be extracted for coupling to heliospheric software), the former interpretation is preferred. This means that when deriving the physical interpretation of the simulation results, it will be assumed that the inner boundary is at the distance of 2$\;R_\oplus$.

\subsection{Convergence}

 \begin{figure*}
   \centering
   \includegraphics[width=15cm]{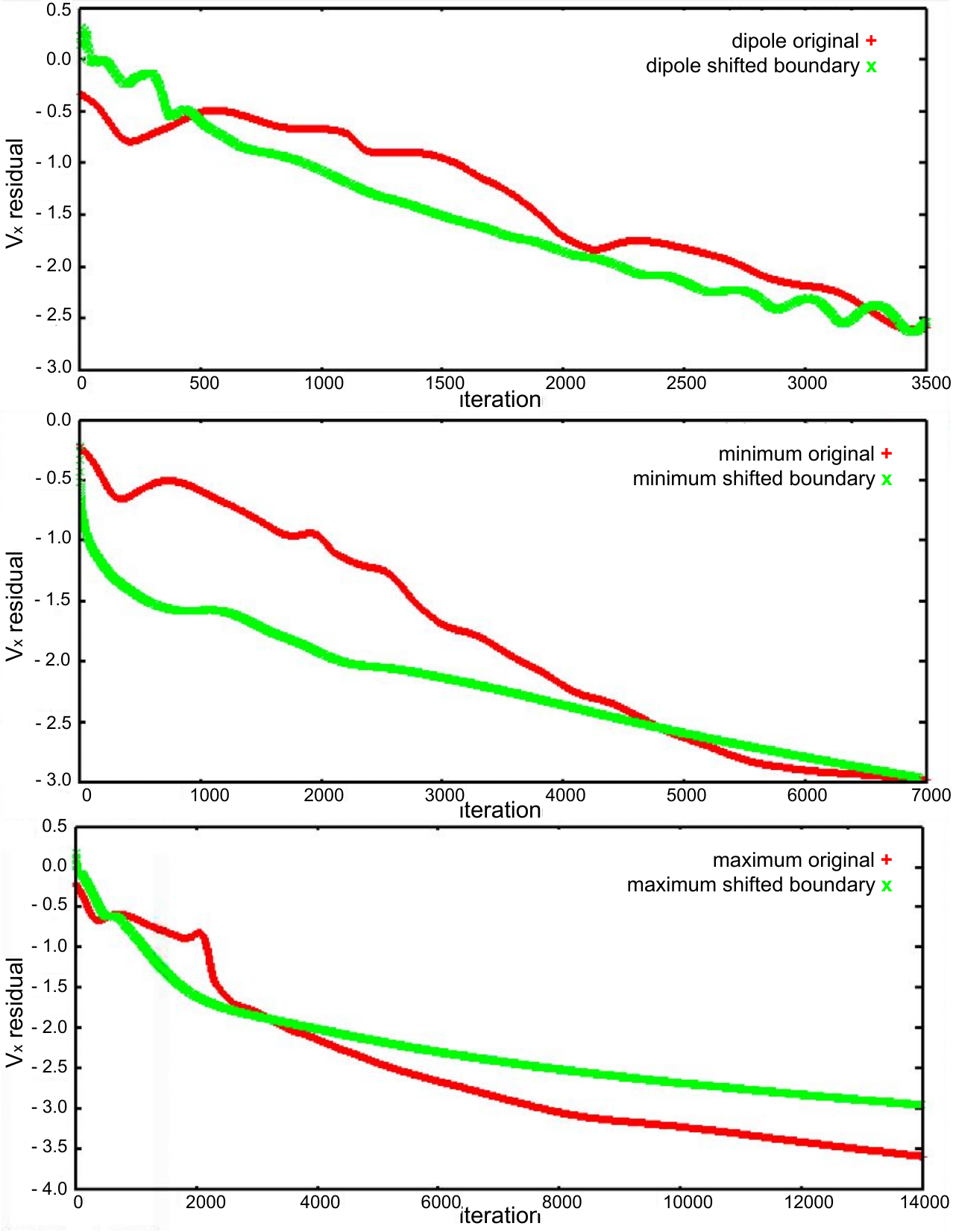}
   \caption{Comparison of convergence based on the $V_x$ residual for the original setup (red) and the new setup (green) in the dipole case (top), minimum solar activity case (middle) and maximum solar activity case (bottom).}
              \label{fig:convergence}%
    \end{figure*}

 \begin{figure*}
   \centering
   \includegraphics[width=15cm]{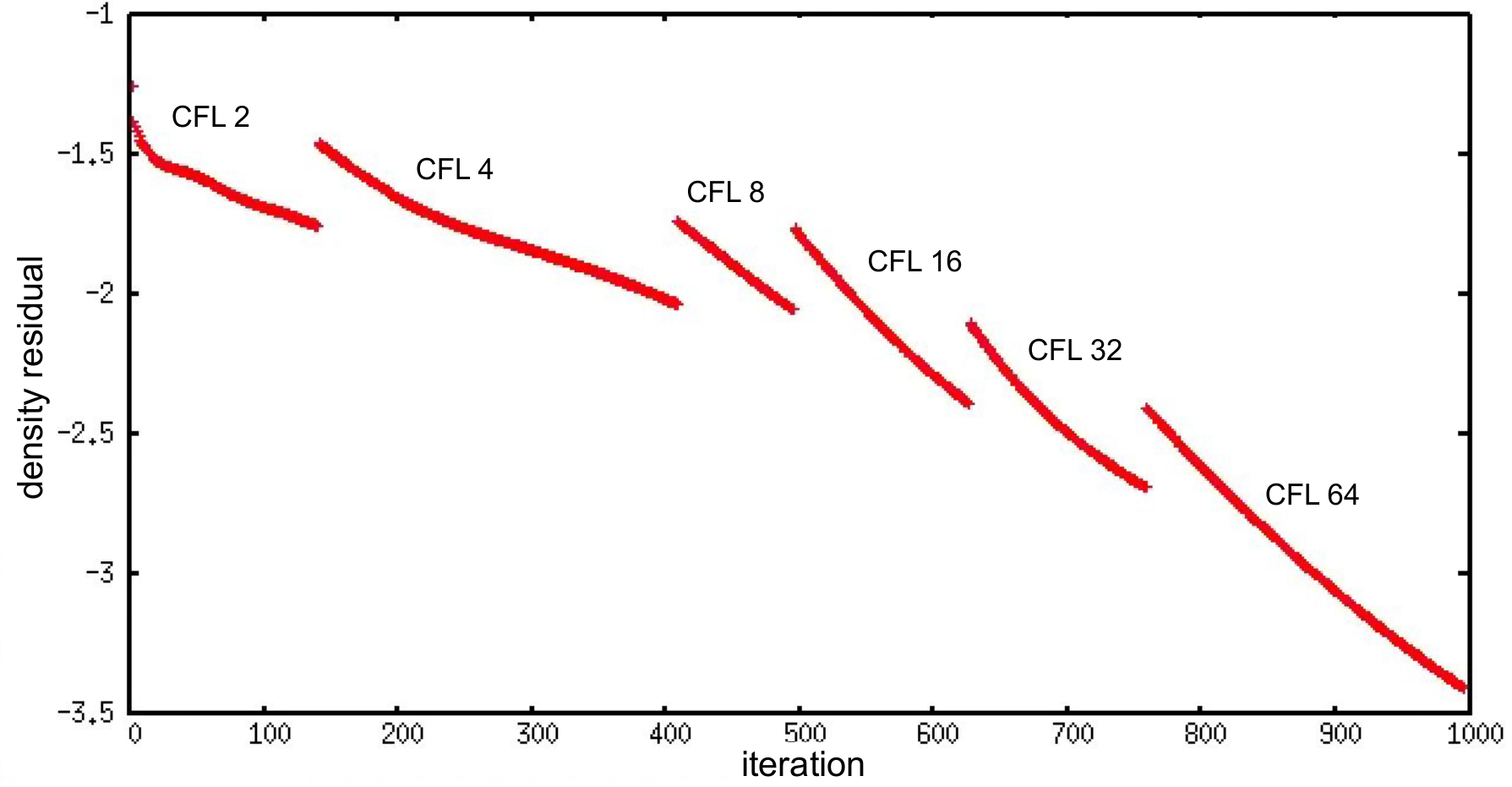}
   \caption{Demonstration that rapid convergence (here $\rho$ residual) with steep CFL profiles in just 1000 iterations is still possible with the shifted-boundary approach.}
              \label{fig:ConvergenceCFL}%
    \end{figure*}
    
Lastly, it is also essential to discuss convergence of the setups, since the primary purpose of the COCONUT solver is to be integrated into operational model chains within VSWMC for space weather forecasting. 

The cases presented above (dipole, 2008 minimum and 2012 maximum) were ran with a constant CFL set to 1, both with the original setting and with the shifted-boundary to 2$\;R_\oplus$ in order to fairly determine how the residual evolves over time and when it reaches the lower pre-set limit. The simulations were ran using the same grid resolution, partitioning, number and type of cores on the Flemish Supercomputer Centre. The residual chosen for monitoring was the $V_x$ residual (the $x$ component of the velocity), since the residual curves show its adimensional absolute value and these value ranges were, unlike density or pressure, similar for both setups. The results, with the original setup in red and shifted-boundary setup in green are shown in Figure~\ref{fig:convergence}, with the top representing the dipole case, the middle the minimum of activity and the bottom the maximum of activity case.

From Figure~\ref{fig:convergence}, it can be concluded that in all the cases, the shifted-boundary setup had a similar convergence performance when compared to the original setup at the same grid. While a constant CFL was held for a fair comparison in the aforementioned Figure, this setup still allows for much steeper CFL profiles with higher CFL values, see Figure~\ref{fig:ConvergenceCFL} where the $\rho$ residual is shown for the 2008 minimum case with a variable CFL run. Here, it is demonstrated that with gradually growing CFL values, a good convergence in just 1000 iterations (and roughly 1 hour of compute time on 144 cores) can be achieved for a realistic case with this method. 

Considering the fact that the level of resolution in the results of the former is far more enhanced, the grid can be further coarsened, which will further improve the convergence time and stability. Furthermore, while for now the same grid was used for a fair comparison of convergence (where it was scaled up for the 2$\;R_\oplus$-boundary case), in operational runs, the outermost layers of the 2$\;R_\oplus$-boundary grid can be removed as they are far beyond 25$\;R_\oplus$, which will lead to further performance enhancement. 

\subsection{Possible challenges of the approach}

It was demonstrated that the newly applied approach leads to significantly more pronounced profiles with more realistic features compared to the original setup. It does, however, pose some limitations on the use of the code, which must also be discussed.

Firstly, clearly in this setting, we cannot accurately study the features close to the solar surface. The 1$\;R_\oplus$ to 2$\;R_\oplus$ region of corona cannot be studied as it is not included in the domain. The region close to the 2$\;R_\oplus$ boundary also cannot be used for accurate analysis for the reasons outlined in \ref{ss:validity} (e.g., the formation of loops or streamers), but this is generally considered unnecessary as this setup is meant for operational runs, not for academic research purposes. Beforehand, coronal structures were validated using solar eclipse data (derived for 1$\;R_\oplus$). With this approach, coronagraph data starting from 2$\;R_\oplus$ to 2.5$\;R_\oplus$ could be used for that purpose. 

Secondly, another potential obstacle that will have to be overcome with this setup is the fact that the majority of coronal heating term formulations to achieve bimodal wind have been developed considering the lower boundary to be at 1$\;R_\oplus$, with the parameters and proportionality constants adjusted in kind. If we do, in future, switch to a 2$\;R_\oplus$ inner boundary formulation, the formulation of the heating terms will have to be adjusted to produce similar results, possibly complicating the inclusion of these terms in the first place. 

On the other hand, the lower solar atmosphere is highly dynamical as it is strongly affected by the turbulent convection, and this is difficult for a global coronal model to properly describe. Therefore, shifting the inner boundary could avoid the corresponding numerical and physical difficulties. 

As for now, the demonstrated benefits appear to overcome these potential drawbacks for our polytropic MHD model.

\section{Conclusions}
\label{sec:conclusion}

In this paper, after having shown the effects of reducing the magnetic field strength and gradients during magnetogram pre-processing on the resolution of features in the COCONUT global coronal model simulations, we have proposed a simple and original method to counterbalance this simplification without losing computational performance. 

After a short presentation of the spherical harmonics magnetogram pre-processing and literature review, we have concluded that the magnetic fields which we generally use as an input to the COCONUT model are unrealistically under-resolved. As demonstrated on a minimum solar activity case (2008 total eclipse), this inevitably leads to smaller electromagnetic forces and magnetic pressure with respect to gravitational forces and thermal pressure of the plasma, resulting in an nonphysical plasma $\beta$ values, consequently leading to a lack of electromagnetic structures in density and velocity profiles. Increasing the prescribed magnetic field strength, however, deteriorates the computational performance and robustness of the code.

Therefore, we have formulated an approach to mitigate this issue by placing the inner boundary further away from the star. We find that the respective magnetic fields that we use should be realistic at the height of roughly 1$\;R_\oplus$ from the photosphere, which means that we shift our inner boundary to the distance 2$\;R_\oplus$, lowering the prescribed density and pressure accordingly. 

By applying this new setup on case of a magnetic dipole, a minimum of solar activity (2008 total eclipse) and a maximum of solar activity (2012 total eclipse), we have demonstrated that this approach leads to far more pronounced density profiles, correctly following the magnetic field lines. We have also shown that these new density profiles lead to more physical and better resolved synthetic WLI, now showcasing features despite the fact that the setup still lacks physical terms such as heating terms, wave pressure, radiation and conduction. For the minimum activity case, we have also validated the new technique through comparison with tomography data, and it is revealed that the density profiles resulting from the new approach fit the tomography data much better even at the distance of 4$\;R_\oplus$. 

Next, it was discussed that prescribing a 1$\;R_\oplus$ magnetogram at 2$\;R_\oplus$ might lead to larger features than what is realistic and change the shape of these features as a result. Due to the fact that for space weather forecasting we mostly focus on how these features extend in the domain and not how exactly they look close to the inner boundary, with the former being shown to be much better with the new adopted approach, we argue that despite this weakness, the proposed approach is still a suitable for our purposes. It is also pointed out that due to the non-dimensional nature of the MHD equations, the new approach can be also interpreted as an approach in which we preserve the original domain starting at 1$\;R_\oplus$, but in which we artificially decrease the thermal pressure and the gravity to achieve a realistic plasma $\beta$ at the inner boundary. In that case, the above-mentioned problem with larger features disappears, but the prescribed thermal pressure and gravity are not representative of reality. 

Finally, since COCONUT is meant for operational runs, we have also evaluated its convergence with the modified inner boundary settings. From the behaviour of the velocity $x$-component residual, it is concluded that the convergence performance is similar to the original setup on the same grid. In addition, for the new setup, an even coarser grid can be afforded since it provides results with much better sharpness compared to the original, and with a smaller domain, it also requires fewer cells. Thus, the performance of the solver in terms of the required computational resources will only increase. Using this approach may result in the necessity to adjust our validation means in the future, focusing more on coronagraph data instead of WLI from total solar eclipses. In addition, shifting the inner boundary might make it difficult to apply some of the most popular heating terms available in literature, as these were generally developed for setups starting at 1$\;R_\oplus$. This will be addressed in future work, when implementing these terms into COCONUT. All in all, since this technique demonstrated superior performance and accuracy, it is worthwhile to consider it for operational runs despite these projected possible difficulties. 

\begin{acknowledgements}
This work has been granted by the AFOSR basic research initiative project FA9550-18-1-0093. 
This project has also received funding from the European Union’s Horizon 2020 research and innovation program under grant agreement No.~870405 (EUHFORIA 2.0) and the ESA project "Heliospheric modelling techniques“ (Contract No. 4000133080/20/NL/CRS).
These results were also obtained in the framework of the projects
C14/19/089  (C1 project Internal Funds KU Leuven), G.0B58.23N  (FWO-Vlaanderen), SIDC Data Exploitation (ESA Prodex-12), and Belspo project B2/191/P1/SWiM.
The resources and services used in this work were provided by the VSC (Flemish Supercomputer Centre), funded by the Research Foundation - Flanders (FWO) and the Flemish Government.
Wilcox Solar Observatory data used in this study was obtained via the web site \url{http://wso.stanford.edu} courtesy of J.T.\ Hoeksema.
The Wilcox Solar Observatory is currently supported by NASA.
Data were acquired by GONG instruments operated by NISP/NSO/AURA/NSF with contribution from NOAA.
HMI data are courtesy of the Joint Science Operations Center (JSOC) Science Data Processing team at Stanford University.
This work utilises data produced collaboratively between AFRL/ADAPT and NSO/NISP. \\
Data used in this study was obtained from the following websites: \\
WSO: \url{http://wso.stanford.edu/synopticl.html} \\
GONG: \url{https://gong2.nso.edu/archive/patch.pl?menutype=z} \\
HMI: \url{http://jsoc.stanford.edu/HMI/LOS_Synoptic_charts.html} \\
GONG-ADAPT: \url{https://gong.nso.edu/adapt/maps/}
\end{acknowledgements}

%
%

\bibliographystyle{aa} 
\bibliography{biblio.bib}

\end{document}